\documentclass[submission,copyright,creativecommons]{eptcs}
\providecommand{\event}{QAPL 2017} 
\usepackage{underscore}

\usepackage[utf8]{inputenc}

\usepackage{mathtools}
\usepackage{amssymb}
\usepackage{amsthm}
\usepackage{wrapfig}
\usepackage{multirow,bigdelim}
\usepackage{color}
\usepackage{graphicx,tikz}
\usepackage{caption}
\usepackage{subcaption}

\DeclareMathOperator{\Act}{Act}
\DeclareMathOperator{\Pre}{Pr}

\DeclareMathOperator{\Var}{Var}

\theoremstyle{definition}
\newtheorem{definition}{Definition}
\newtheorem{example}{Example}

\newcommand{\qedd}{\hfill\ensuremath{\diamond}}

\title{SEA-PARAM:\\
Exploring Schedulers in Parametric MDPs}
\author{Sebastian Arming
\institute{University of Salzburg, Austria}
\and
Ezio Bartocci
\institute{TU Wien, Austria}
\and
Ana Sokolova
\institute{University of Salzburg, Austria}
}
\def\titlerunning{SEA-PARAM}
\def\authorrunning{S. Arming, E. Bartocci, and A. Sokolova}
\begin{document}
\maketitle

\begin{abstract}
We study parametric Markov decision processes (PMDPs) and their 
reachability probabilities "independent" of the parameters. Different 
to existing work on parameter synthesis (implemented in the tools 
PARAM and PRISM), our main  focus is on describing different types 
of optimal deterministic memoryless schedulers for the whole parameter 
range. We implement a  simple prototype tool SEA-PARAM that computes 
these optimal schedulers and show experimental results. 
\end{abstract}

\section{Introduction}\label{sec:intro}

A Markov decision process (MDP)~\cite{Baier2008} is a state-based 
Markov model in which a state can perform one of the available action-labeled 
transitions after which it ends up in a next state according to a probability 
distribution on states. The choice of a transition to take is nondeterministic, 
but once a transition is chosen the behaviour is probabilistic.
 MDPs provide a valuable mathematical framework to solve control and dependability problems in a wide range of applications, 
 including the control of epidemic processes~\cite{lefevre81}, power management~\cite{Qui2001}, queueing systems~\cite{Sennott1998}, and cyber-physical systems~\cite{Ayala2012}. 
 MDPs are also known as reactive probabilistic systems~\cite{LS91:ic,GlabbeekSS95} and closely related to probabilistic automata~\cite{SL94:concur}.

In this paper, we study \emph{parametric} Markov decision processes (PMDPs)~\cite{ChenHHKQ013,HahnHZ11}. These are models in which (some of) the transition probabilities depend on a set of parameters. An example of an action in a PMDP is tossing a (possibly unfair) coin which lands heads with probability $p$ and tails with probability $1-p$ where $p \in [0,1]$ is a parameter. Hence, a PMDP represents a whole family of MDPs---one for each valuation of the parameters.

We study reachability properties in PMDPs. To explain what we do exactly, let us take a step back. If an MDP can only perform a single action in each state, then it is a Markov chain (MC). If a PMDP can perform a single action in each state, then it is a \emph{parametric} Markov chain (PMC). Given a start state and a target state in a PMC, the probability of reaching the target from the start state is a \emph{rational function} in the set of parameters. This rational function can be elegantly computed by the method of Daws~\cite{DC05} providing arithmetic interpretation for regular expressions. The method has been further developed and efficiently implemented in the tool PARAM~\cite{HHWZ2010,HahnHZ11b,HahnHZ11}.

 Clearly, there is no such thing as the probability of reaching a target state from a starting state in an MDP: such a reachability probability depends on which actions were taken along the way, i.e. of how the nondeterministic choices were resolved. What is usually of interest though are the min/max reachability probabilities, i.e. among all possible ways to resolve the nondeterministic choices, those that provide minimal/maximal probability of reaching a state. Nondeterministic choices are resolved using schedulers or policies, and luckily when it comes to min/max reachability probabilities \emph{simple} schedulers suffice~\cite{Baier2008}. Simple schedulers are deterministic and memoryless, i.e. history independent. Given an MDP, a simple scheduler induces an MC, and the reachability probabilities under this scheduler are simply the reachability probabilities of the induced MC. 

With PMDPs, the situation is even more delicate. The probability of reaching a target state from a starting state depends on the scheduler, i.e. on how the nondeterministic choices were resolved, as well as on the values of the parameters. The full reachability picture looks like a sea --- each scheduler imposes a rational function --- a \emph{wave} --- over the parameter range; the sea then consists of all the waves. 

There are two possible scenarios of interest:
\begin{itemize}
\item[(1)] We have access to the parameters.
\item[(2)] We have no access to the parameters, they represent uncertainty or noise or choices of the environment.
\end{itemize}

In case (1), parameter synthesis is the problem to solve. The parameter synthesis problem comes in two flavours: (a)~Find the parameter values that maximise / minimise the reachability probability; (b)~For each value of the parameters, find the max/min reachability probability. These are the problems that have attracted most attention in the analysis of PMCs~\cite{BartocciGKRS11,JansenCVWAKB14,PathakAJTK15,
 DehnertJJCVBKA15,QuatmannD0JK16} and PMDPs~\cite{ChenHHKQ013,HahnHZ11}, see Section~\ref{sec:rel-work} for more details.

In this paper, we consider case (2) and propose solutions for imposing bounds on the reachability probabilities throughout the whole parameter range. 

In particular we:
\begin{itemize}
	\item Start by enumerating all simple schedulers and computing their corresponding rational functions. 
	\item Identify classes of optimal schedulers, for different problems of interest, see Section~\ref{sec:scheds}. 
	\item Provide a tool that computes optimal schedulers in each of the classes for a given PMDP, see Section~\ref{sec:imp-exp}.
\end{itemize} 
    
  We admit that we take upon this task knowing that it is computationally hard. Already the number of simple schedulers is exponential in the number of states of the involved PMDP. Optimisation is in general also hard (computing maxima, minima, and integrals of the involved rational functions), see Section~\ref{sec:imp-exp} for references and more details. Nevertheless, the analysis that we aim at is not an online analysis, but rather a preprocessing step, and even if it may only work on small examples, it provides insight in the behaviour of a system and its schedulers. 
    
  Our tool extensively uses the state-of-the-art tools PARAM1~\cite{HHWZ2010,HahnHZ11b} and PARAM2~\cite{HahnHZ11} for efficient computation of the rational functions, see Section~\ref{sec:rel-work} and Section~\ref{sec:imp-exp} for details. Once we have all schedulers and their respective waves, we feed the waves to a numerical tool that allows us to calculate the optimal schedulers.

\begin{wrapfigure}{R}{0.3\textwidth}
\includegraphics[width=0.25\textwidth]{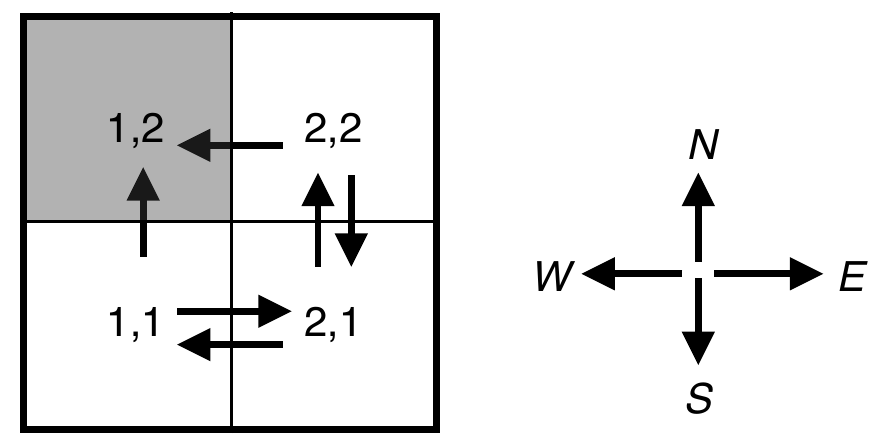}
\caption{$2\times 2$ Labyrinth with sink at (1,2)}	
\label{fig:2x2pic-ex}
\end{wrapfigure}

We experiment with a class of examples describing the behaviour of a robot walking in a labyrinth grid. Each position on the grid is a state of the MDP, and the available actions are $N$, $S$, $E$, and $W$, describing the directions (north, south, east, and west) of a possible move. Not all actions are available in every state. Some states represent holes (sinks) in which no action is available, others correspond to border-positions, and hence some actions are disabled.
See Section~\ref{sec:examples} for further description of our class of examples. One small concrete example in this class is the PMDP describing a $2\times 2$ grid with a sink at position $(1,2)$.    In $(1,1)$ the actions $N$ and $E$ are available, in $(2,1)$ the actions $N$ and $W$, and in $(2,2)$, the actions $S$ and $W$.

\begin{figure}\centering
\begin{subfigure}{0.5\textwidth}
\centering
\includegraphics[width=0.45\linewidth]{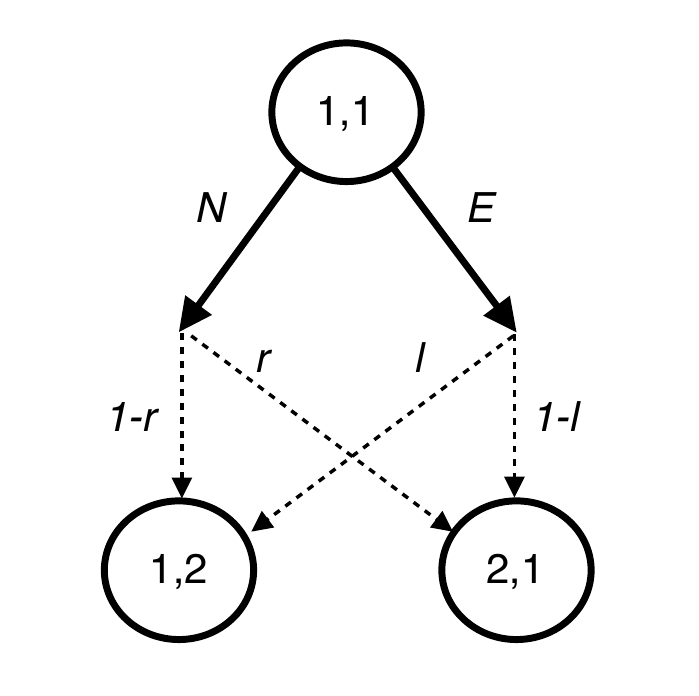}
\caption{Fixed failure}	
\label{fig:state11-ff-ex}
\end{subfigure}%
\begin{subfigure}{0.5\textwidth}
\centering
\includegraphics[width=0.45\linewidth]{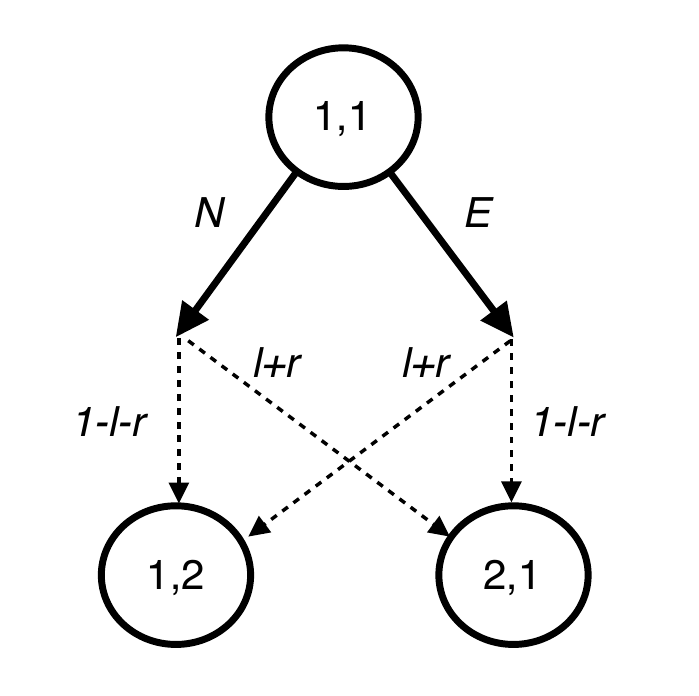}
\caption{Fixed success}	
\label{fig:state11-fs-ex}
\end{subfigure}
\caption{The behaviour of state $(1,1)$}	
\label{fig:state11-ex}
\end{figure}

\begin{figure}\centering
\begin{subfigure}{0.5\textwidth}
\centering
\includegraphics[width=0.45\linewidth]{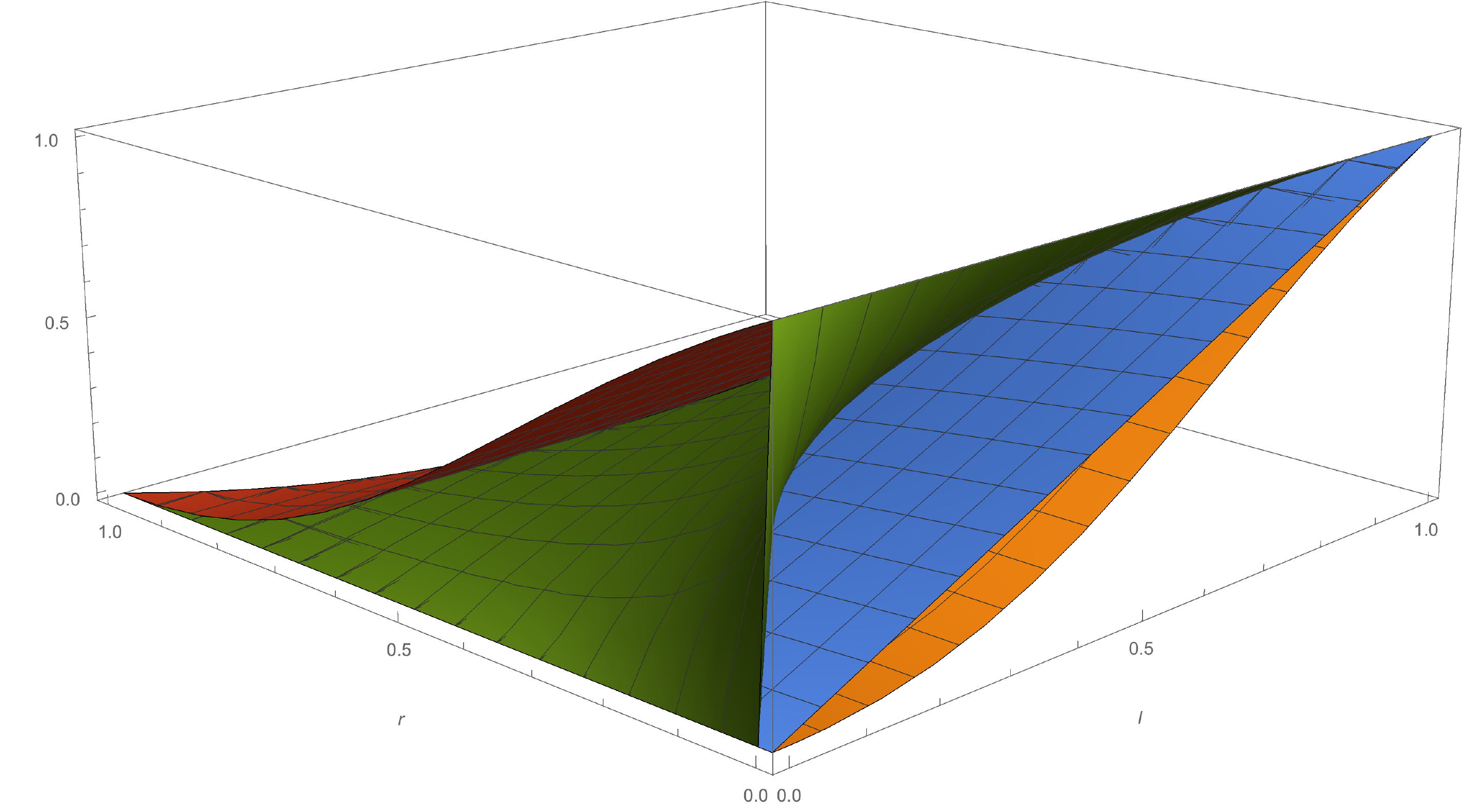}
\caption{Fixed failure}	
\label{fig:2x2-sea-ff}
\end{subfigure}%
\begin{subfigure}{0.5\textwidth}
\centering
\includegraphics[width=0.45\linewidth]{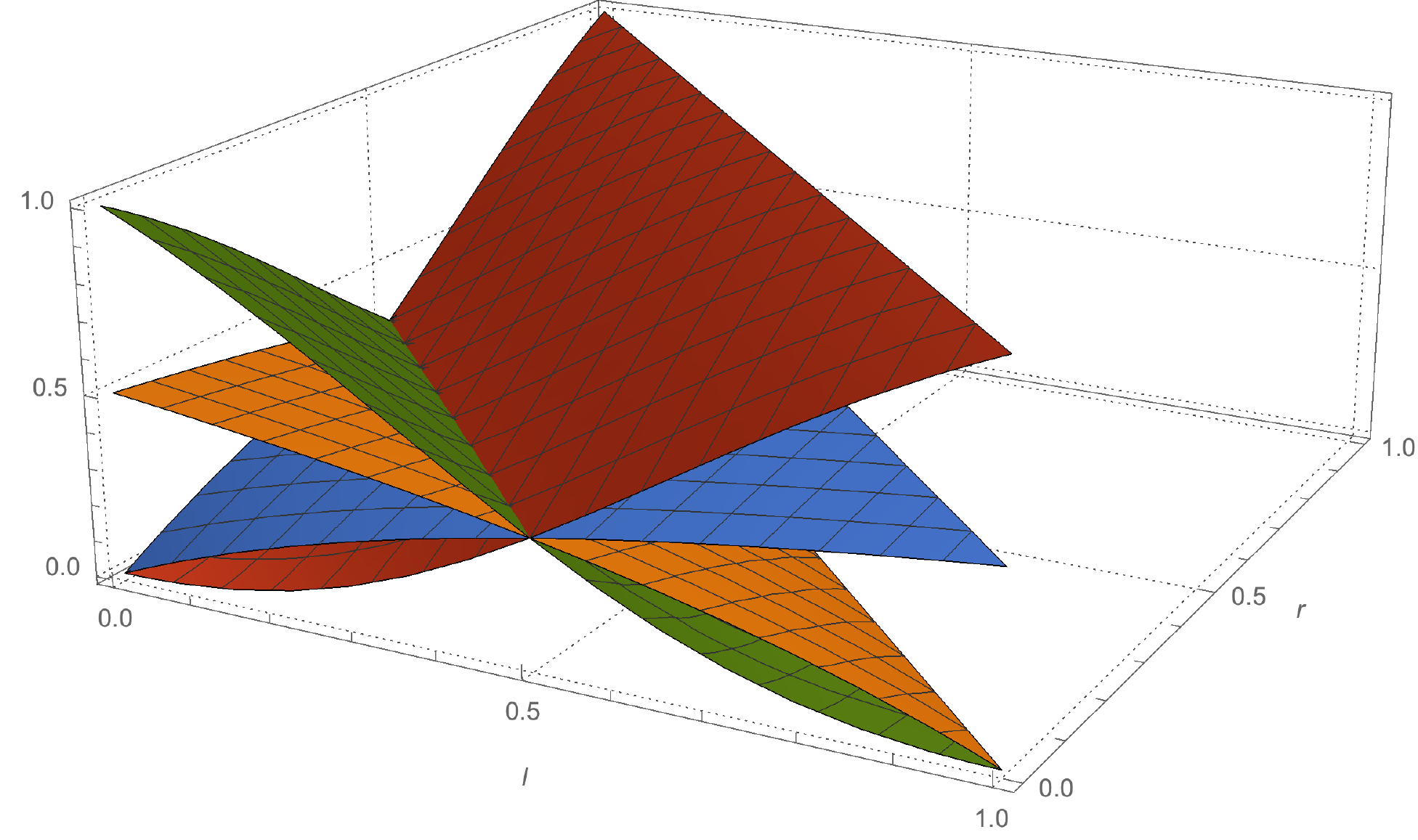}
\caption{Fixed success}	
\label{fig:2x2-sea-fs}
\end{subfigure}
\caption{The sea of reachability probabilities from state $(1,1)$ to state $(2,2)$}	
\label{fig:2x2-sea}
\end{figure}

The model is parametric with two parameters $l$ and $r$ having the following meaning: In a state $s$ with an enabled action $M$, the robot moves forward with probability $1-l-r$ to its intended state $s'$, or ends up in the state $s_l$ left of $s$ (in the direction of the move) with probability $l$, and in the state $s_r$ right of $s$ (in the directions of the move) with probability $r$, provided both states $s_l$ and $s_r$ exist. If one of $s_l$ or $s_r$ does not exist, then we consider two scenarios:
\begin{itemize}
	\item Fixed failure: In this scenario, the probability to the existing state $s_l$ or $s_r$ remains $l$ or $r$, but the probability of reaching $s'$ increases to $1-l$ or $1-r$, respectively.
	\item Fixed success: Here, the probability to $s'$ remains the same,  and the probability to $s_l$ or $s_r$ (whichever exists) becomes $l+r$.
\end{itemize} 
Figure~\ref{fig:state11-ex} shows the behaviour of state $(1,1)$ in both scenarios and
Figure~\ref{fig:2x2-sea} pictures all waves -- rational functions corresponding to reachability of target state $(2,2)$ from the starting state $(1,1)$ -- in each of the two scenarios. 

As we can see even in this small example, the reachability probability varies through the parameter range and significantly depends on the chosen scheduler. For different purposes, different schedulers may be preferred. We identify ten classes of optimal schedulers that may be preferred in certain cases. For example, one may wish to use a scheduler that guarantees highest reachability probability for any value of the parameter, if such a scheduler exists. We call such a scheduler \emph{dominant}. That would be the red scheduler in Figure~\ref{fig:2x2-sea-ff}. In Figure~\ref{fig:2x2-sea-fs} there is no dominant scheduler. However, one may prefer a scheduler that reaches the maximum value (reachability probability $1$ in this case) for some value of the parameter. We call such schedulers \emph{optimistic}. In  Figure~\ref{fig:2x2-sea-fs} the red and the green schedulers are optimistic. In addition, one may prefer the red over the green, as under the assumption of a uniform distribution of parameters, the red has a higher value over a larger parameter region --- we call such schedulers \emph{expectation} schedulers. For yet another purpose, one may prefer the yellow or the blue scheduler, as its difference in reachability probabilities is the smallest --- the \emph{bound} scheduler according to our definition.

See Section~\ref{sec:scheds} for the exact definitions of all classes of optimal schedulers.

Finally, we mention that we started analysing simple schedulers as a first step in the scheduler analysis of PMDPs. As we discuss in Section~\ref{sec:conc}, we are aware that optimal schedulers (in our classes) need not be simple. Nevertheless, we believe that conquering simple schedulers is an important first step.

\section{Related Work}\label{sec:rel-work}

In the last decade there has been a growing interest in studying parametric probabilistic 
models~\cite{DC05,LMST2007} where some of the probabilities (or rates) in the  models 
are not known a-priori.  These models are very useful when certain quantities (e.g. fault rates, 
packet loss ratios, etc.) are partially available (which is often the case) 
or unavailable at the design time of a system.
In his seminal work~\cite{DC05}, Daws studies the problem of symbolic model checking of parametric 
probabilistic Markov chains. He provides a method based on regular expressions extraction and state elimination
to symbolically express the probability to reach a target state from a starting state as a multivariate rational function whose domain is the parameter space. 
 This technique was further investigated and implemented in the PARAM1 and PARAM2  tools~\cite{HHWZ2010,HahnHZ11b} 
 and it is now also included in the popular PRISM model checker~\cite{KwiatkowskaNP11}.  
 In this context, the problem of \emph{parameter synthesis}  for a parametric Markov chain consists
 of solving a constrained  nonlinear optimisation problem where the objective function
 is  a multivariate rational function  representing the probability to satisfy a given reachability property
  depending on the parameters.
As discussed in~\cite{Kreinovich1998bi,LMST2007} and later in this paper, when the order of these multivariate rational functions  
 is high, such constrained optimisation problem can become computationally very expensive.
 
In~\cite{BartocciGKRS11} Bartocci et al. introduce a complementary technique
to \emph{parameter synthesis}, called \emph{model repair},
that exploits the PARAM1 tool in combination with a nonlinear optimisation tool to find automatically
 the minimal change of the parameter values required for a  model to  satisfy
 a given reachability property that the model originally violates.  
 In this case the problem boils down to solving a nonlinear optimisation program having for
 \emph{objective function} an L2-norm (quadratic and indeed suitable for convex optimisation)
 measuring the distance between the original parameter values and the new ones and having
  as  \emph{constrains} the multivariate rational function associated
  with the reachability property.
 
 Recently, more sophisticated symbolic parameter synthesis techniques~\cite{JansenCVWAKB14,PathakAJTK15,
 DehnertJJCVBKA15,QuatmannD0JK16}
 based also on SMT solvers and greedy approaches~\cite{PathakAJTK15} have further improved this
 field of research.
 At the same time statistical-based approaches leveraging powerful machine learning
 techniques~\cite{BortolussiMS16,BartocciBNS15} have been shown to provide better
 scaling of the model checking problem for large parametric continuous Markov chains when 
the number of parameters is limited and the event of satisfying the property is not rare.  

All the aforementioned methods
 do not natively support nondeterministic choice and are indeed not suitable for solving
 parametric Markov decision processes.  The parametric model checking problem
 for this class of models has been addressed so far in the literature using two complementary
 methods~\cite{ChenHHKQ013,HahnHZ11}.
 
 The first method, implemented in PARAM2~\cite{HahnHZ11}, is a
 \emph{region-based approach} where the parameter space is divided into regions
 representing sets of parameter valuations. For each region, lower
and upper bounds on optimal parameter values are computed by
evaluating the edge points of the regions.  Given a desired level
of precision for the result as input, the algorithm decides whether to further split the region
into smaller ones to be explored or to terminate with the intervals found. 
The correctness and the termination of this algorithm is guaranteed 
only under certain assumptions as discussed in~\cite{HahnHZ11}.
The second method~\cite{ChenHHKQ013} is a sampling-based approach 
(i.e. based on sampling methods like the Metropolis-Hastings algorithm,
particle swarm optimisation, and the cross-entropy method) that are used 
to search the parameter space.  These heuristics usually do not guarantee 
that global optimal parameters will be found. Furthermore, when the regions of the parameters satisfying 
a requirement are very small, a large amount of simulations is required.

We just became aware  of a very recent work of Cubuktepe et al.~\cite{Cubuktepe2017} 
(to appear in TACAS'17) where the authors consider the problem 
of parameter synthesis in parametric Markov decision processes 
using signomial programs, a class of nonconvex optimisation problems
for which it is possible to provide suboptimal solutions.

\section{Markov Chains and Markov Decision Processes}\label{sec:mdp}

\begin{definition}[Markov chain]\label{def:MC}
A (discrete-time) Markov chain (MC) is a pair
$M = (S, P)$
where:
\begin{itemize}
\item
    $S$ is a countable set of states, and
\item
    $P\colon S\times {S} \rightarrow [0,1]$ is a transition
    probability function such that for all $s$ in $S$, $\sum_{s'\in{S}}P(s,s')=1$.\qedd
\end{itemize} 
\end{definition}

Given an MC $M = (S,P)$ and two states $s,t\in S$, we denote the
probability to reach $t$ from $s$ by $\Pre^M(s,t)$. If $M$ is clear from the context, we will omit the superscript in the reachability probability.

We next present the definition of an MDP without atomic propositions and rewards, as they do not play a role for what follows.

\begin{definition}[Markov decision process]\label{def:MDP}
A (discrete-time) Markov Decision Process (MDP) is a triple
$M = (S, \Act,  P) $
where:
\begin{itemize}
\item
$S$ is a countable set of states,
\item
$\Act$ is a set of actions,
\item 
$P\colon S\times \Act \times{S} \rightarrow [0,1]$ is a transition probability 
 function such that for all $s$ in $S$ and $a$ in $\Act$ we have $\sum_{s'\in{S}}P(s,a, s') \in \{0,1\} $.\qedd
\end{itemize}
\end{definition}

In this paper we only consider finite MCs and MDPs, that is MCs  and MDPs in which the set of states $S$ (and actions $\Act$) is finite.
If needed, we may also specify a distinguished initial state $s_0 \in S$ in an MC or an MDP.

An action $a$ is \emph{enabled} in an MDP state $s$ iff $\sum_{s'\in{S}}P(s,a, s') = 1$. 
We denote by $\Act(s)$ the set of enabled actions in state $s$. It is often required that $\Act(s) \neq \emptyset $ in an MDP, but we omit this requirement. A state $s$ for which $\Act(s) = \emptyset$ is called a \emph{sink}.
A \emph{simple scheduler} resolves the  nondeterministic choice, 
selecting at each non-sink state $s$ one of the enabled actions $a \in \Act(s)$. A synonym for a simple scheduler is deterministic memoryless/history-independent scheduler.  

\begin{definition}[Simple scheduler] \label{def:scheduler}
Given an MDP $M = (S, \Act, P)$, a {simple scheduler} $\xi$ of $M$ 
 is a function $\xi \colon S \to \Act + 1$ where $1 = \{\bot\}$ and $+$ denotes disjoint union, satisfying 
$\xi(s) \in \Act(s)$ for all $s \in S$ such that $\Act(s) \neq \emptyset$, and $\xi(s) = \bot$ otherwise.\qedd
\end{definition}
 
\begin{definition}[Scheduler-induced Markov chain]\label{def:IMC}
Let $\xi$ be a simple scheduler of an MDP $M$. Then the $\xi$-{induced Markov chain} is the Markov chain
$M_\xi = (S,P_\xi)$ where
$P_\xi(s,t) = P(s,\xi(s),t)$ if $\xi(s) \neq \bot$ and $P_\xi(s,s) = 1$ otherwise.\qedd
\end{definition}

Note that in this work we only consider simple schedulers. This justifies our nonstandard (and much simpler) definition of an induced Markov chain.
From now on we will sometimes simply say \emph{scheduler} for a simple scheduler. 

\begin{definition}[Maximum/Minimum reachability probabilities] \label{def:min-max-reach-prob}
Given an MDP $M = (S,\Act,P)$ and two states $s,t\in S$, the
maximum reachability probability from $s$ to $t$ is
    \[
        \Pre_{\max}^M(s,t) = \max_\xi \Pre^{M_\xi}(s,t),
    \]
\noindent
and similarly, the minimum reachability probability from $s$ to $t$ is given by
    \[
        \Pre_{\min}^M(s,t) = \min_\xi \Pre^{M_\xi}(s,t),
    \]
\noindent where $\xi$ ranges over all simple schedulers. We call a scheduler $\xi$ a \emph{maximal (minimal) scheduler from $s$
to $t$}
iff $\Pre^{M_\xi}(s,t)$ is the maximal (minimal) reachability probability from $s$ to $t$.\qedd
\end{definition}

\section{Parametric MCs and MDPs} \label{sec:pmdp}

We first recall the notion of a rational function (following~\cite{HHWZ2010,HahnHZ11b}, with a small restriction). Let $V = \{x_1, \dots, x_n\}$ be a fixed set of variables. An \emph{evaluation} is a function $v \colon V \to \mathbb{R}$. A \emph{polynomial} over $V$ is a function
$$g(x_1, \dots,x_n) = \sum_{i_1, \dots, i_n} a_{i_1, \dots, i_n}x_1^{i_1} \cdots x_n^{i_n},$$
where $i_j \in \mathbb{N}$ for $1 \le j \le n$ and each $a_{i_1, \dots, i_n} \in \mathbb{R}$. A \emph{rational function} over $V$ is a quotient $$f(x_1, \dots, x_n) = \frac{g_1(x_1,\dots,x_n)}{g_2(x_1,\dots,x_n)}$$ of two polynomials $g_1$ and $g_2$ over $V$. By $\mathcal{F}_V$ we denote the set of rational functions over $V$. Hence, a rational function is a symbolic representation of a function from $\mathbb{R}^n$ to $\mathbb{R}$. Given $f \in \mathcal{F}_V$ and an evaluation $v$, we write $f\langle v\rangle$ for $f(v(x_1), \dots, v(x_n))$.   

It is now straightforward to extend MCs and MDPs with parameters~\cite{DC05,LMST2007,HHWZ2010,HahnHZ11b}. Again, we only consider finite models.

\begin{definition}[Parametric Markov chain]\label{def:PMC}
A parametric (discrete-time) Markov chain (PMC) is a triple $M = (S, V, P) $
where:
\begin{itemize}
\item
    $S$ is a finite set of states, 
\item $V$ is a finite set of parameters, and
\item $P:S\times {S} \rightarrow \mathcal{F}_V$ is  the parametric probability transition function.\qedd
\end{itemize}
\end{definition}

Given a PMC $M = (S,V,P)$, a valuation $v$ of the parameters induces an MC $M_v = (S, P_v)$ where $P_v(s,s') = P(s,s')\langle v\rangle$ for all $s,s' \in S$, if for all $s$ in $S$ we have $\sum_{s'\in{S}}P(s,s')\langle v\rangle=1$. If a valuation $v$ induces a Markov chain on $M$, then we call $v$ \emph{admissible}. The set of all admissible valuations for $M$ is the \emph{parameter space} of $M$.

Similarly, we define parametric MDPs.

\begin{definition}[Parametric Markov Decision Process]\label{def:PMDP}
A parametric (discrete-time) Markov Decision Process (PMDP) is a tuple
$M = (S, \Act, V,  P) $
where:
\begin{itemize}
\item
$S$ is a finite set of states,
\item
$\Act$ is a finite set of actions,
\item
$V$ is a finite set of parameters, and
\item 
$P:S\times \Act \times{S} \rightarrow \mathcal{F}_V$ is the parametric transition probability function.\qedd
\end{itemize}
\end{definition}

Also here a valuation may induce an MDP from a PMDP, in which case we call it \emph{admissible}.
Given a PMDP $M = (S,\Act,V,P)$, a valuation $v$ of the parameters induces an MDP $M_v = (S,\Act, P_v)$ where $P_v(s,a,s') = P(s,a,s')\langle v\rangle$, if for all $s$ in $S$ and $a$ in $\Act$ we have $\sum_{s'\in{S}}P(s,a,s')\langle v\rangle\in \{0,1\}$. 
Also here, the set of admissible valuations is the \emph{parameter space} of $M$.

Notice that a PMDP $M$ and its $v$-induced MDP $M_v$ have the same set of states and actions, as well as the same sets of enabled actions in each state, and therefore they have the same simple schedulers. Now, starting from a PMDP $M$, and given its scheduler $\xi$, one may: (1) first consider the $\xi$-induced PMC $M_\xi$ and then the $v$-induced MC $({M_\xi})_v$ for a valuation $v$, or (2) one first takes the valuation-induced MDP $M_v$ and then its scheduler-induced MC $({M_v})_\xi$. The result is the same and hence we write $M_{\xi,v}$ for the $\xi$-and-$v$-induced MC. 

We now fix a source state $s$ in a PMDP, and a target state $t$ and discuss the reachability probabilities that are now dependent on both the choice of a scheduler $\xi$ and the choice of a parameter valuation $v$. Given a valuation $v$ and a scheduler $\xi$, the reachability probability is $\Pre^{M_{\xi,v}}(s,t)$. The (reachability probability) \emph{wave} corresponding to $\xi$ is a rational function $f_\xi$ in the set of parameters, such that $f_\xi\langle v\rangle = \Pre^{M_{\xi,v}}(s,t)$. The (reachability probability) \emph{sea} consists of all $f_\xi$ for all schedulers $\xi$. 

We also write (for a PMDP $M$):
$$\begin{array}{lcl}
        \Pre_{\max}^{M_v}(s,t) & = & \max_\xi \Pre^{M_{\xi,v}}(s,t),\\
        \Pre_{\max}^{M_\xi}(s,t) & = & \max_v \Pre^{M_{\xi,v}}(s,t),\\
        \Pre_{\max}^{M}(s,t) & = & \max_{\xi,v} \Pre^{M_{\xi,v}}(s,t),\\
\end{array}$$
and similarly for the minimum reachability probabilities.

\section{Classes of Optimal Schedulers} \label{sec:scheds}

In this section we define and discuss a selection of types of optimal schedulers. This is meant to serve as an invitation for the reader to further develop useful notions of optimality. 

Our initial idea is the following: Once we have generated all rational functions (corresponding to all schedulers), a type of optimality assigns a score to each rational function (and hence to the scheduler inducing it). The optimal schedulers of this type then maximise or minimise the assigned score.

We introduce the notion of a \emph{dominant scheduler} and nine additional types of optimal schedulers. These types are:
the optimistic, the pessimistic, the bound, the expectation, the stable, the $\varepsilon$-bounded, the $\varepsilon$-stable, and the $\varepsilon$-bounded- and $\varepsilon$-stable-robust.
We next present the definition for each of them. For simplicity, we may use scheduler and function interchangeably --- thus identifying a scheduler and its induced rational function when no confusion may arise.

\begin{definition}[Dominant scheduler] \label{def:dominant}
A scheduler $\omega$ is dominant if at any parameter valuation $v$, its function has the maximal value of all functions of all schedulers, i.e.
$\forall v. \forall \xi. f_\omega\langle v\rangle \ge f_\xi\langle v\rangle.$ \qedd
\end{definition}

\begin{definition}[Optimistic scheduler] \label{def:optimistic}
A scheduler $\omega$ is optimistic, if its function has the maximal maximum value of all functions of all schedulers, i.e.\\ \hspace*{3cm} $\Pre_{\max}^{M_\omega}(s,t) = \max_\xi \Pre_{\max}^{M_\xi}(s,t) = \Pre_{\max}^{M}(s,t).$\qedd
\end{definition}

\begin{definition}[Pessimistic scheduler] \label{def:pessimistic}
A scheduler is pessimistic, if its function has the maximal minimum value of all functions of all schedulers, i.e. \\ \hspace*{3cm} $\Pre_{\min}^{M_\omega}(s,t) = \max_\xi \Pre_{\min}^{M_\xi}(s,t).$\qedd
\end{definition}

\begin{definition}  [Bound scheduler]\label{def:bound}
A scheduler is bound, if its function has the minimal range, i.e. minimal difference between its maximal and  minimal value of all functions of all schedulers, i.e. \\ \hspace*{3cm} $\Pre_{\max}^{M_\omega}(s,t) - \Pre_{\min}^{M_\omega}(s,t) = \min_\xi \left(\Pre_{\max}^{M_\xi}(s,t) - \Pre_{\min}^{M_\xi}(s,t)\right).$\qedd
\end{definition}

\begin{definition}  [$\varepsilon$-Bounded scheduler]\label{def:eps-bounded}
A scheduler $\xi$ is $\varepsilon$-bounded if the length of the (closed-interval) range of its function is bounded by $\varepsilon$, i.e. \\ \hspace*{3cm} $\Pre_{\max}^{M_\xi}(s,t) - \Pre_{\min}^{M_\xi}(s,t) \le \varepsilon$\\
\noindent for a non-negative real number $\varepsilon$.\qedd
\end{definition}

\begin{definition} [$\varepsilon$-Bounded robust scheduler]\label{def:eps-bounded-robust}
A scheduler $\omega$ is $\varepsilon$-bounded robust if it is the maximal among all $\varepsilon$-bounded schedulers, i.e. $\forall v. \forall \textrm{~$\varepsilon$-bound~}\xi. f_\omega\langle v\rangle \ge f_\xi\langle v\rangle.$\qedd
\end{definition}

The intuition behind these types of optimal schedulers is the following. If a user does not know the value of the parameters, then taking the 
\begin{itemize}
\item dominant scheduler guarantees that one can do as good as it gets independent of the parameters;
\item optimistic scheduler guarantees that one can do as good as it gets in case the parameters are the best possible; 
\item pessimistic scheduler guarantees that no matter what the parameters are, even in the worst case we will perform better than the worst case of any other scheduler;
\item bound scheduler guarantees that one will see minimal difference in reachability probability by varying the parameters;
\item $\varepsilon$-boundness is an absolute notion guaranteeing that such a scheduler never has a larger difference in reachability probability than $\varepsilon$;
\item finally, $\varepsilon$-bounded robustness gives the maximal scheduler among all $\varepsilon$-bounded ones.
\end{itemize}

Dominant, $\varepsilon$-bounded, and $\varepsilon$-bounded robust schedulers need not exist. 

Note that computing optimistic, pessimistic, bound, $\varepsilon$-bounded, and $\varepsilon$-bounded robust schedulers requires computing the maximum and the minimum of the involved functions, which is in general hard~\cite{Kreinovich1998bi}, see Section~\ref{sec:imp-exp} for more details. 

The following classes do not require computing extremal values and may provide a better global picture of the reachability probabilities. Their optimality is based on maximising/minimising or bounding the probability mass over the whole parameter space, also allowing for specifying a probability distribution on the parameter space.  If the distribution of parameters is unknown, we assume uniform distribution. However, it is likely that a distribution of parameters is known or can be estimated, in which case these schedulers take it into account.  From now on, Let $p$ denote a probability density function over the parameter space.

Before we proceed, let us define the expectation and variance of a scheduler. The expectation of a scheduler $\xi$ is $E(\xi) = E(f_\xi) = \int f_\xi\, dp$ and the variance is $\Var(\xi) = E(\xi - E(\xi))^2$. Note that here $\xi - E(\xi)$ denotes the rational function $f_\xi - E(\xi)$.

\begin{definition}[Expectation Scheduler] \label{def:expected}
A scheduler is an expectation scheduler, if its function has the maximal expected value of all functions of all schedulers, i.e.  $\omega$ is an expectation scheduler if $E(\omega) = \max_\xi E(\xi)$.\qedd
\end{definition}

\begin{definition} [Stable scheduler]\label{def:stable}
A scheduler $\omega$ is stable, if its function has the minimal variance, i.e.  \\ \hspace*{3cm} $\Var(\omega) = \min_\xi \Var(\xi).$\qedd
\end{definition}

\begin{definition} [$\varepsilon$-Stable scheduler]\label{def:eps-stable}
A scheduler $\xi$ is $\varepsilon$-stable if its variance is bounded by $\varepsilon$, i.e. \\ \hspace*{3cm} $\Var(\xi) \le \varepsilon$\\
\noindent for a non-negative real number $\varepsilon$.\qedd
\end{definition}

\begin{definition} [ $\varepsilon$-Stable robust scheduler]\label{def:eps-stable-robust}
A scheduler $\omega$ is $\varepsilon$-stable robust if it its expectation is maximal among all $\varepsilon$-stable schedulers, i.e. $ E(\omega) = \max_{\textrm{~$\varepsilon$-stable~}\xi} E(\xi).$\qedd
\end{definition}

\noindent If a dominant scheduler exists, then it is also optimistic, pessimistic, and expectation optimal.

\begin{example}
Consider the $2\times 2$ labyrinth with sink at $(1,2)$ from Figure~\ref{fig:2x2pic-ex} in the introduction.
 
In the fixed failure case, Figure~\ref{fig:2x2-sea-ff}, the red scheduler is dominant (and hence optimistic, pessimistic, and expectation optimal). All schedulers are optimistic, pessimistic, and bound. The yellow scheduler is stable, and the blue is (median variance)-stable.

In the fixed success case, Figure~\ref{fig:2x2-sea-fs}, there is no dominant scheduler. The red and green schedulers are optimistic, all are pessimistic, the yellow and the blue are bound. The red is expectation optimal, the yellow is stable, and the blue is (median variance)-stable robust.  	
\end{example}

\section{Parametric Labyrinths}\label{sec:examples}

The class of examples of a robot in a labyrinth provides a wide playground for studying parametric models. We consider $n \times n$ labyrinths. States are the positions in the labyrinth, and the set of actions is $\{N, S, E, W\}$. 

Taking an action probabilistically determines the next state, as the robot may indeed reach the intended new position or fail to do so and end up in another unintended position. There are many ways to specify what happens if the robot fails, we chose the way as in the example in the introduction: our robot can fail to reach the intended position and instead end up left or right of its current position with a certain probability. 

A most general way to turn this into a parametric model is to consider all probabilities depending on a parameter, e.g. in every state, for every action, there is a parameter that provides the probability to fail left and another that provides the probability to fail right, and the probability of success is determined by the values of these two parameters. This results in a model with $8|S| = 8n^2$ parameters. 

We simplify this general scenario and limit the parameters to smaller numbers. In particular, we consider models with $k$ parameters where
\begin{itemize}
\item[(1)] $k = 8$ and we take per action two parameters (e.g. for action $N$, the probability to fail left with action $N$ and the probability to fail right), which are then the same in every state whenever this action is taken.
\item[(2)] $k=2$ and we take two parameters $l$ and $r$ that serve the purpose like in (1) and in the example in the introduction for every state and every action.
\item[(3)] $k=1$ and we have a single parameter $p$ in the model that serves the purpose like in (2) for every state and every action. 
\end{itemize}

In all of these cases for states on the boundary we consider one of the two scenarios -- fixed failure or fixed success -- as specified for the example in the introduction. 

In addition, we experiment with making some states sink states, just like we did with state $(1,2)$ in the introduction example.

\section{Implementation and Experiments}\label{sec:imp-exp}

We have implemented a first prototype of SEA-PARAM leveraging 
the open-source parametric model checking framework of the PRISM 
model checker~\cite{KwiatkowskaNP11} and Wolfram Mathematica\footnote{https://www.wolfram.com/mathematica/}.

SEA-PARAM receives as input a PMDP and a reachability property.
Firstly, it explores all the possible memoryless schedulers generating 
for each of them a multivariate rational function that maps 
the parameter space into the probability to satisfy the 
desired property.   For the generation 
and the manipulation of the multivariate rational functions, 
PRISM leverages the Java Algebra Systems (JAS)\footnote{http://krum.rz.uni-mannheim.de/jas/}.
This task is embarrassingly parallel,
since each memoryless scheduler can be 
treated independently  from the others. We exploit this with a concurrent implementation, which leads to constant (given by the number of cores) speed-up.  
However, in the worst-case the number of schedulers (which we straightforwardly enumerate in this first attempt)
can be exponential in the number of states, resulting in exponential running time. 

After the memoryless schedulers enumeration and function computation, 
the corresponding multivariate rational functions are 
evaluated  according to a chosen optimality criterion
using a script developed within the Wolfram Mathematica 
framework.  We chose Mathematica for the ability to quickly 
implement our different formal 
notions of optimality criteria for the schedulers provided in the paper. 
The Mathematica program takes as input the list of schedulers with their corresponding functions 
generated in the previous step and computes a score 
for each multivariate rational function.  
This task  can again be computed in parallel for each 
multivariate rational function.
Nevertheless, again, in general the computation of the score of a multivariate rational function is NP-hard~\cite{Kawamura:2011tx,DBLP:journals/siamcomp/Sahni74,DBLP:journals/mp/MurtyK87}.
For example, already the minimisation of a multi-variate quadratic function
over the unit cube is NP-hard, see e.g.~\cite{DBLP:journals/mp/MurtyK87} for a
reduction from SUBSET-SUM.
For several classes of well-behaved functions (e.g.~convex functions
or unimodal ones) our scores can be efficiently computed.
We know for sure that not all our functions are convex or unimodal, but there
is still a chance that the functions form another well-behaved class.
We intend to explore this possibility in future work.

Note that, since we generate a list of schedulers together with their rational functions, it is straightforward to find the scheduler corresponding to a rational function.

\subsection{Experiments}\label{sec:exp}

\begin{table}[t]
  \begin{tabular}{|lllll|rr|rrr|}
    \hline
    \multicolumn{5}{|c|}{Grid scenario} & \multicolumn{2}{c|}{Number of} & \multicolumn{3}{c|}{Execution time in seconds} \\
k & size & type & target & sinks & schedulers & functions & PRISM & optimistic & expectation \\ \hline
8 & 2x2 & ff & (2,2) & (1,2) & 4 & 4 & 0.11 & 1.39 & 1.40 \\
8 & 2x2 & fs & (2,2) & (1,2) & 4 & 4 & 0.10 & 2.19 & 19.09 \\ \hline
2 & 2x2 & ff & (2,2) & (1,2) & 4 & 4 & 0.07 & 0.58 & 0.71 \\
2 & 2x2 & fs & (2,2) & (1,2) & 216 & 63 & 1.76 & 1.90 & 0.26 \\
2 & 3x3 & ff & (1,3) & (1,2),(2,2) & 432 & 120 & 5.68 & 3.94 & 0.69 \\
2 & 3x3 & ff & (2,2) & (1,2) & 864 & 398 & 13.53 & 15.36 & 2.11 \\
2 & 3x3 & ff & (3,3) & (1,2) & 648 & 246 & 6.95 & 7.83 & 0.98 \\
2 & 3x3 & ff & (3,3) & (2,2) & 4 & 4 & 0.06 & 0.31 & 0.05 \\
2 & 3x3 & fs & (1,3) & (1,2), (2,2) & 216 & 63 & 1.85 & 1.85 & 0.42 \\
2 & 3x3 & fs & (2,2) & (1,2) & 432 & 120 & 6.53 & 4.29 & 0.78 \\
2 & 3x3 & fs & (3,3) & (1,2) & 864 & 399 & 12.14 & 18.64 & 2.63 \\
2 & 3x3 & fs & (3,3) & (2,2) & 648 & 234 & 9.18 & 8.13 & 1.32 \\ \hline
1 & 2x2 & ff & (2,2) & (1,2) & 4 & 4 & 0.06 & 1.40 & 0.04 \\
1 & 2x2 & fs & (2,2) & (1,2) & 216 & 60 & 0.78 & 5.67 & 0.16 \\
1 & 3x3 & ff & (1,3) & (1,2), (2,2) & 432 & 114 & 2.29 & 9.72 & 0.18 \\
1 & 3x3 & ff & (2,2) & (1,2) & 864 & 390 & 6.46 & 34.48 & 0.52 \\
1 & 3x3 & ff & (3,3) & (1,2) & 648 & 122 & 3.48 & 12.57 & 0.19 \\
1 & 3x3 & ff & (3,3) & (2,2) & 4 & 4 & 0.05 & 1.41 & 0.04 \\
1 & 3x3 & fs & (1,3) & (1,2), (2,2) & 216 & 63 & 1.09 & 4.87 & 0.12 \\
1 & 3x3 & fs & (2,2) & (1,2) & 432 & 114 & 1.72 & 11.21 & 0.19 \\
1 & 3x3 & fs & (3,3) & (1,2) & 864 & 391 & 5.63 & 34.01 & 0.54 \\
1 & 3x3 & fs & (3,3) & (2,2) & 648 & 124 & 2.42 & 10.21 & 0.17 \\
1 & 4x4 & fs & (4,4) & (1,2) & 4478976 & 2010270 & 89677.00 & 42824.00 & 32986.00 \\\hline
  \end{tabular}
  \caption{Overview of the experimental results
}
\label{tab:exp}
\end{table}

The experiments reported here ran on a unified memory architecture (UMA)
machine with four 10-core 2GHz Intel Xeon E7-4850 processors supporting two
hardware threads (hyper-threads) per core, 128GB of main memory, and Linux
kernel version 4.4.0.
The first part (based on PRISM SVN revision 11807) was compiled and run with
OpenJDK 1.8.0. The second part was executed in Mathematica 11.0 using 16 parallel kernels.
During our experiments we identified a bug in a greatest-common-divisor (gcd) procedure of the JAS library, that resulted in computing wrong functions.
We work around this bug by substituting a simpler gcd procedure.
All of our code and detailed results of the experiments can be found at
\cite{seaparam}.

Table \ref{tab:exp} gives an overview of our experimental results.
We present 23 experiments in total, for the various scenarios described in Section~\ref{sec:examples}.
For each scenario the table shows the number of schedulers, the number of unique functions (as two schedulers might have the same rational function), as well as the running times of key parts of our system.
In particular, we show the running time for the computation of the rational functions (column PRISM) and the computation of the expectation and optimistic optimal schedulers.
We selected these two classes of optimal schedulers as they illustrate the characteristics of our two score classes (integral/mass vs extremal values) the best.
Our largest experiments involves a 4x4 labyrinth with a single parameter; it
results in over 2 million distinct rational functions (and takes significant
amount of time to compute).

\begin{figure}\centering
  \begin{subfigure}[t]{0.5\textwidth}
    \includegraphics[height=.43\linewidth]{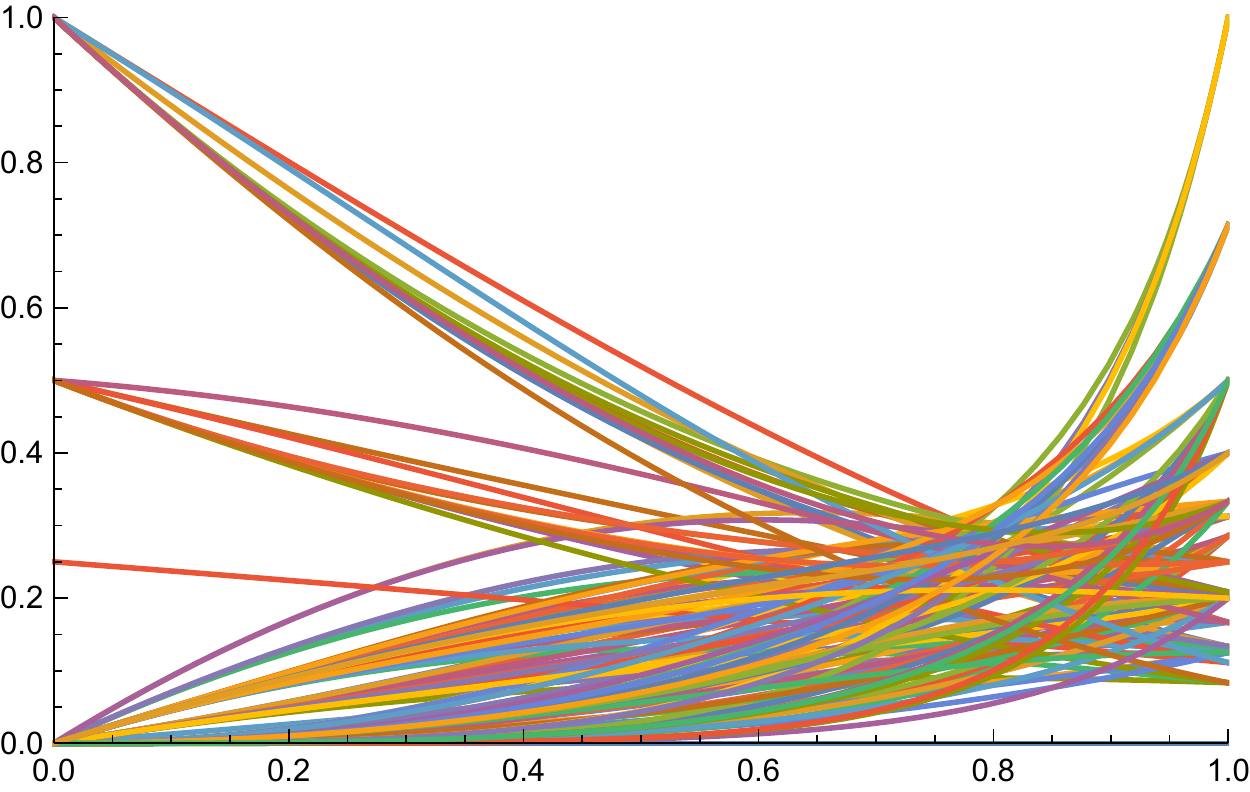}
    \includegraphics[height=.43\linewidth]{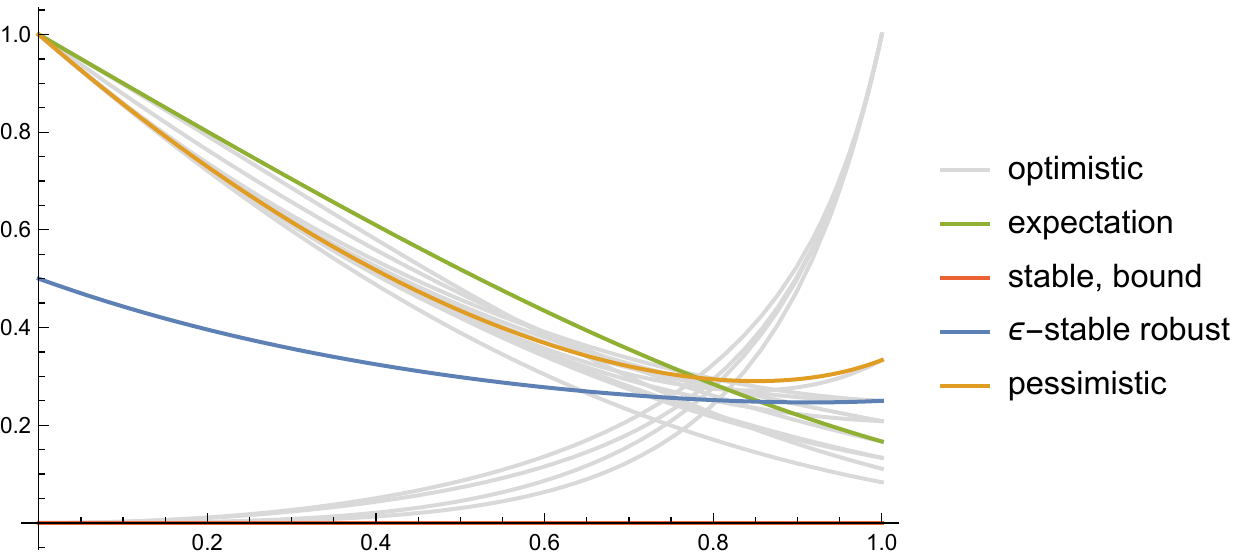}
    \caption{fixed failure from (1,1) to (3,3) with a sink at (2,2)}
    \label{fig:plot-ex-ff}
  \end{subfigure}%
  \begin{subfigure}[t]{0.5\textwidth}
    \includegraphics[height=.43\linewidth]{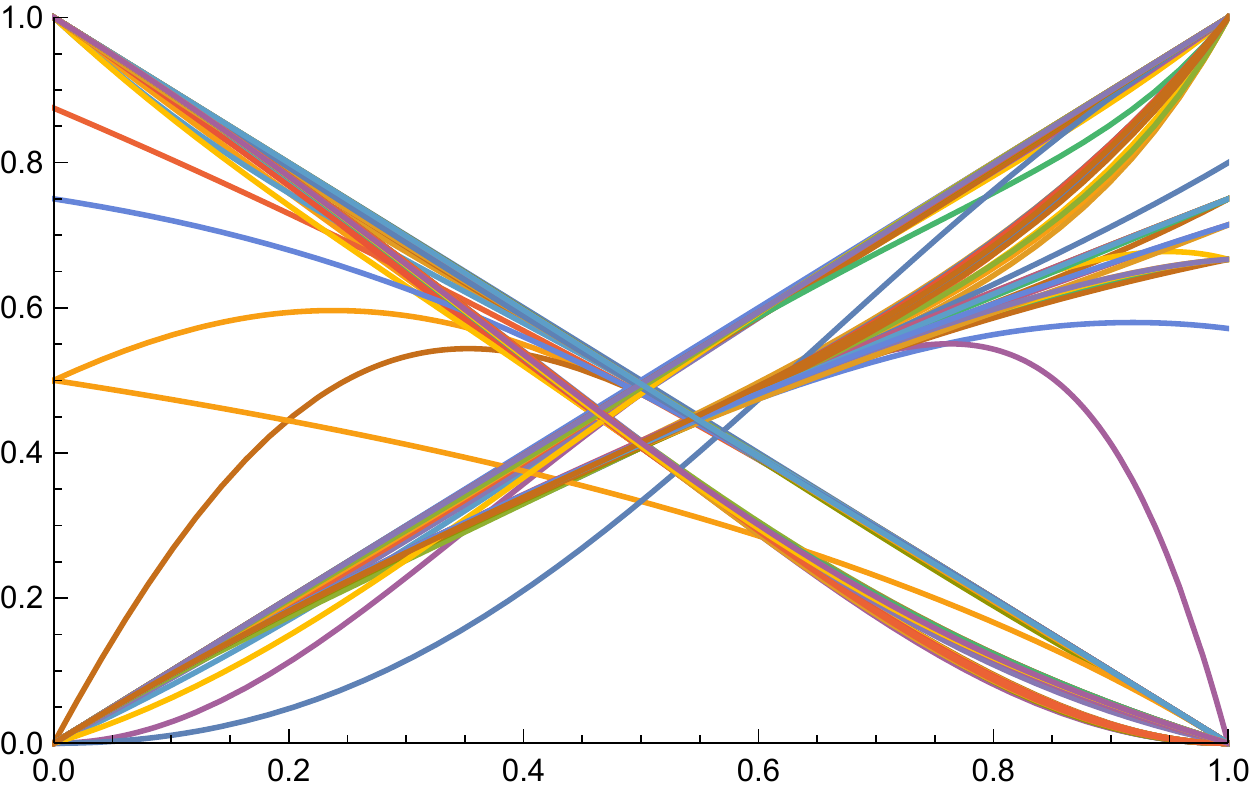}
    \includegraphics[height=.43\linewidth]{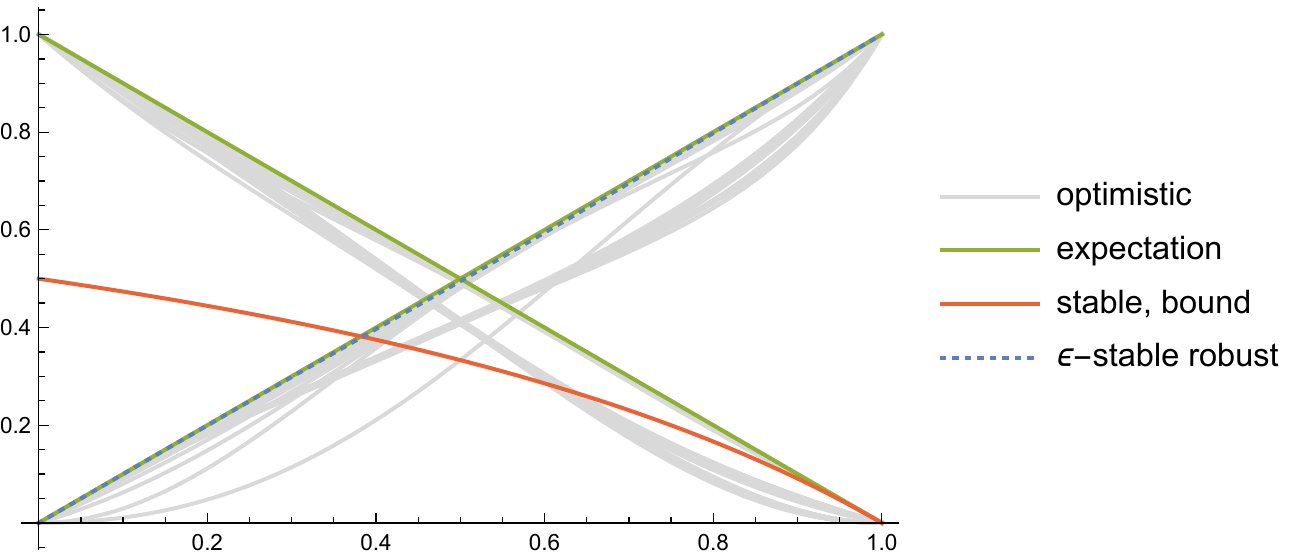}
    \caption{fixed success from (1,1) to (2,2) with a sink at (1,2)}
    \label{fig:plot-ex-fs}
  \end{subfigure}
  \caption{The rational functions of schedulers for two 3x3 labyrinths with $k=1$}
  \label{fig:plot-ex}
\end{figure}

In Figure~\ref{fig:plot-ex} we plot the rational functions of two 3x3 labyrinths with one parameter, to give a flavour of the different schedulers encountered.
In both cases no scheduler is dominant and several of them are optimistic.
In Figure~\ref{fig:plot-ex-ff} we see a single expectation scheduler (actually two schedulers with a single rational function) and a single pessimistic scheduler (again actually two schedulers), while in Figure~\ref{fig:plot-ex-fs} there are two symmetric expectation schedulers, and \emph{all} schedulers are pessimistic (as they all have minimal value $0$).
In both scenarios the stable and the bound schedulers coincide - in Figure~\ref{fig:plot-ex-ff} the corresponding function is constant $0$.
We also show an $\varepsilon$-stable robust scheduler with $\varepsilon$ chosen to be the median variance of the rational functions.
In Figure~\ref{fig:plot-ex-fs} this yields a function very close to the expectation scheduler function with slightly lower variance.

Note that we plot the rational functions for the optimal schedulers.
From these functions we can look up the corresponding schedulers.
For instance, the function labeled expectation in Figure~\ref{fig:plot-ex-ff}
corresponds to the two expectation optimal schedulers that in $(1,1)$ take $E$ or $N$ respectively; take
$N$ in $(1,2)$, $(3,1)$, and $(3,2)$; and take $E$ in $(1,3)$, $(2,3)$ (and of course in $(2,1)$ where there is no other choice).

\section{Discussion}\label{sec:conc}

In our first-version prototype implementation of SEA-PARAM we focus on simple schedulers, which are already exponentially many. However, not always simple schedulers are optimal according to our optimality definitions. A history dependent (hence not simple) scheduler may estimate the parameters and thus provide better behaviour than any simple scheduler, as we show with the following example. 

\begin{figure}[h]\centering
\begin{subfigure}{0.5\textwidth}
\centering
\includegraphics[width=0.8\linewidth]{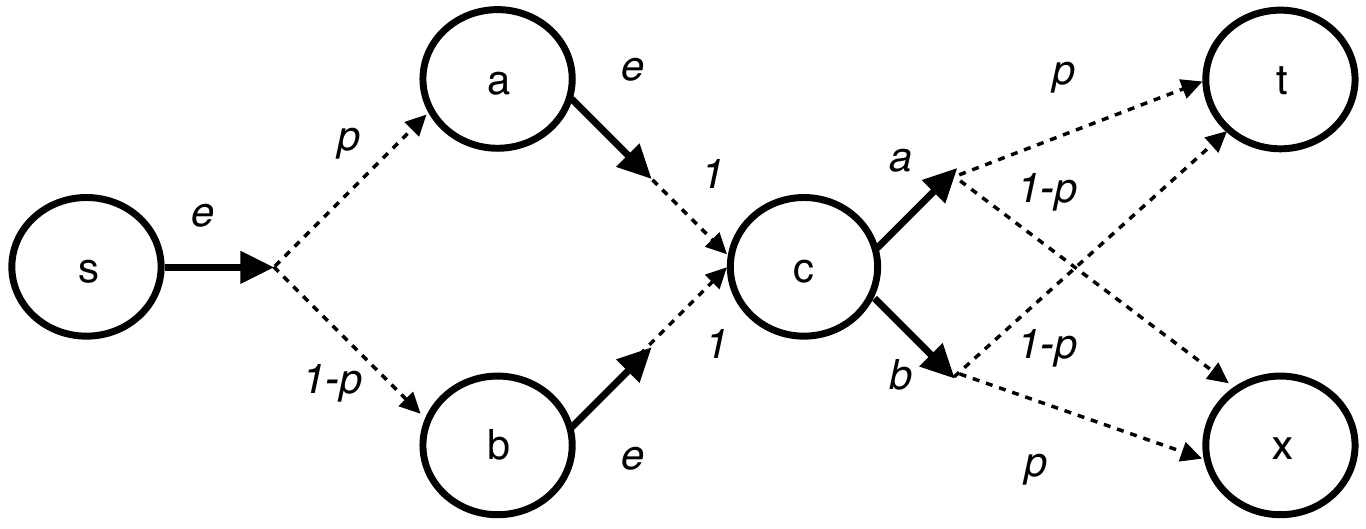}
\caption{An example PMDP $M$}	
\label{fig:history-dependent}
\end{subfigure}
\begin{subfigure}{0.5\textwidth}
\centering
\includegraphics[width=0.95\linewidth]{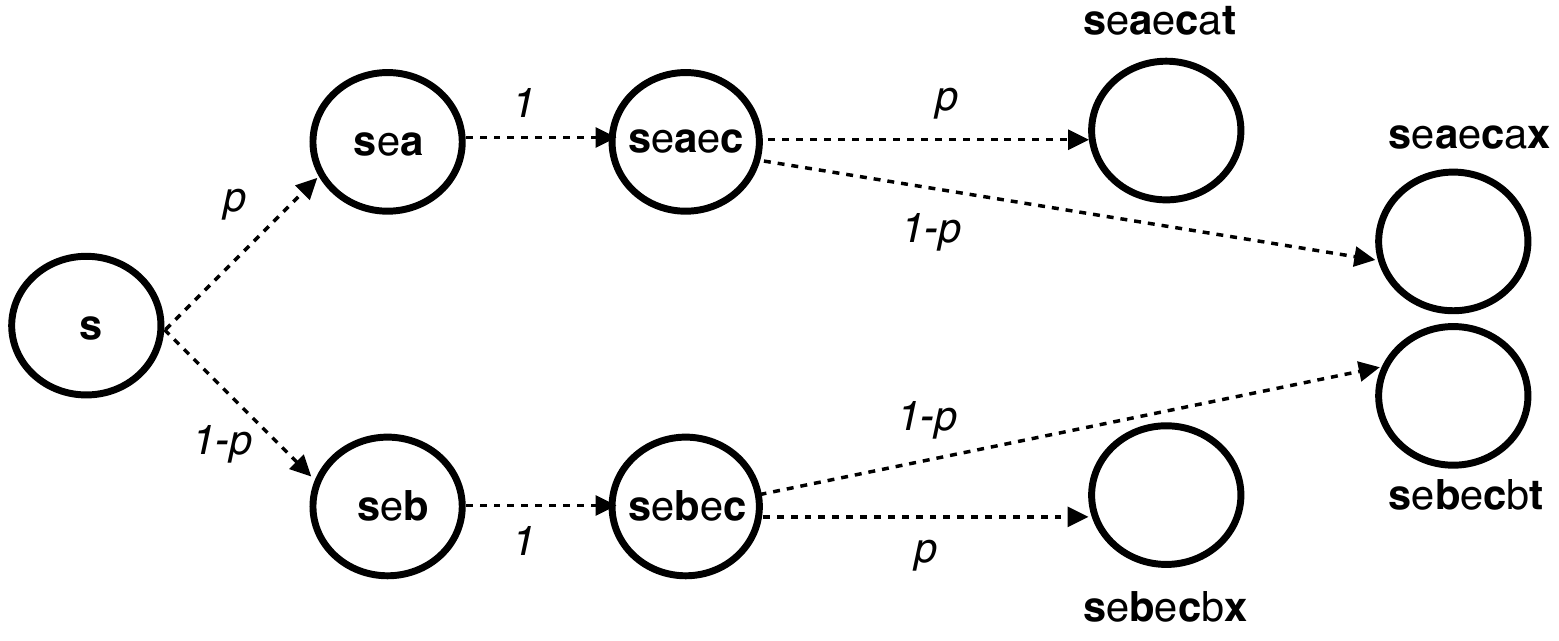}
\caption{Induced MC by a history-dependent scheduler}	
\label{fig:hist-dep-ind-MC}
\end{subfigure}
\caption{History dependency}	
\label{fig:simple-are-not-enough}
\end{figure}

Consider the PMDP $M$ in Figure~\ref{fig:history-dependent} where $p$ is a parameter. There are two simple schedulers for $M$: $\alpha$ with $\alpha(\mathbf{c}) = a$ and $\beta$ with $\beta(\mathbf{c}) = b$ (all other states are mapped to the single available action and $\mathbf{t}$ and $\mathbf{x}$ are sink states). Their corresponding rational functions are $f_\alpha(p) = p$ and $f_\beta(p) = 1-p$. 

\begin{figure}[h]
\centering
\includegraphics[scale=0.8]{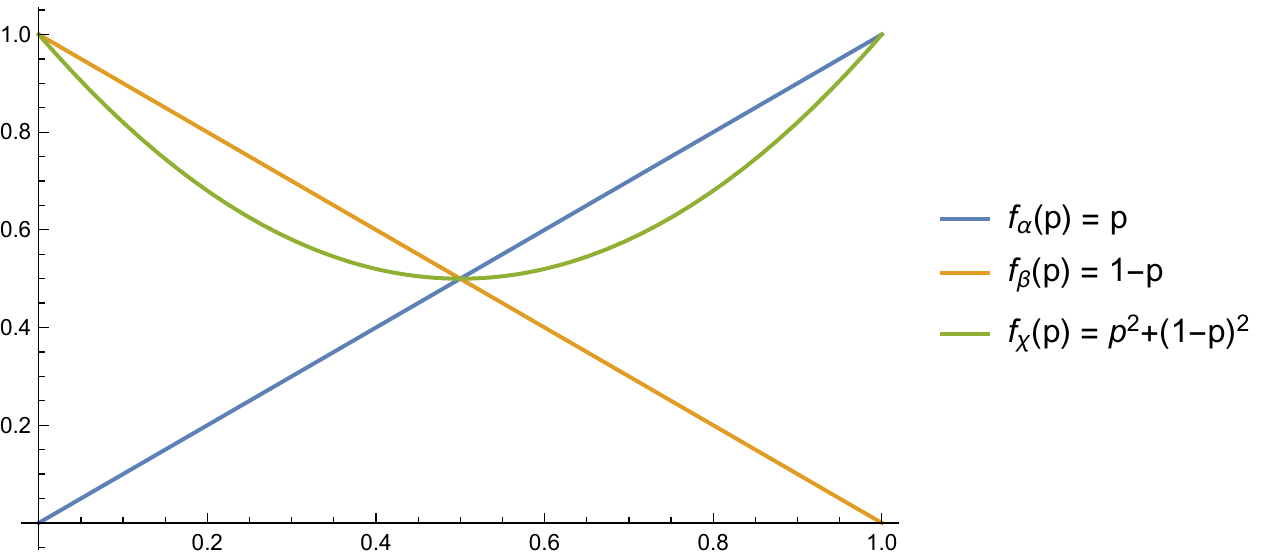}
\caption{The rational functions of $\alpha$, $\beta$, and $\chi$; $\chi$ wins in optimality}	
\label{fig:simple-and-hist-dep}
\end{figure}

Consider now the history-dependent scheduler $\chi$ of $M$ that schedules $a$ in state $\mathbf{c}$ if and only if the state $\mathbf{a}$ has been visited before. The $\chi$-induced MC is shown in Figure~\ref{fig:hist-dep-ind-MC}. The rational function corresponding to $\chi$ is $f_\chi(p) = p^2 + (1-p)^2$.  All three rational functions are depicted in Figure~\ref{fig:simple-and-hist-dep}, and $\chi$ wins in all optimality classes against $\alpha$ and $\beta$. 

We aim at broadening our scheduler exploration to history-dependent schedulers in the near future. 

\vspace*{3mm}
\noindent{}{\bf Acknowledgments.}
This work was supported by the Austrian National Research Network 
RiSE/SHiNE (S11405-N23 and S11411-N23) project funded by the Austrian Science Fund (FWF) 
and partially by the Fclose (Federated Cloud Security) project funded by UnivPM.

\bibliographystyle{eptcs}
\bibliography{related/qapl.bib}
\end{document}